\documentclass[pra,twocolumn,showpacs,amssymb]{revtex4}
\usepackage{graphicx}
\usepackage{amsmath}

\begin{document}

\title{Hofstadter-Hubbard model with opposite magnetic fields:\\
Bardeen-Cooper-Schrieffer pairing and superfluidity in the nearly flat butterfly bands}
   
\author{M. Iskin}
\affiliation{Department of Physics, Ko\c{c} University, Rumelifeneri Yolu, 
34450 Sar\i yer, Istanbul, Turkey}

\date{\today}

\begin{abstract}

Despite the multi-band spectrum of the widely-known Hofstadter butterfly, 
it turns out that the pairing correlations of the time-reversal-symmetric 
Hofstadter-Hubbard model are well-described by a single order 
parameter that is uniform in real space. By exploiting a BCS mean-field 
theory for the nearly-flat butterfly-bands regime of low 
magnetic-flux limit, here we reveal a number of unusual superfluid 
properties both in the ground state and at finite temperatures. 
Our thorough analysis includes but is not limited to the order parameter, 
condensate and superfluid fractions, and the critical BCS and BKT 
transition temperatures.

\end{abstract}


\maketitle

\section{Introduction}
\label{sec:intro}

The Hofstadter model~\cite{hofstadter76} consists only of a tight-binding 
kinetic term on a square lattice, where a spinless quantum particle is 
allowed to tunnel through nearest-neighbor sites with a hopping amplitude 
$t > 0$, and meanwhile it gains an Aharonov-Bohm phase as a 
reflection of the magnetic vector potential. That is the perpendicular 
magnetic field is taken into account via the well-established minimal 
coupling such that the particle acquires $e^{\mathrm{i} 2\pi \alpha}$ 
after traversing a loop around the unit cell. When the magnetic flux $\alpha$ 
corresponds to a ratio $p/q$ of relatively-prime numbers $p$ and $q$, 
the single-particle energy spectrum $\varepsilon_{\mathbf{k}n}$ 
consists of $q$ bands that are indexed here by $n = 0,1,\dots,q-1$
in the first magnetic Brillouin zone. These sub-bands split from 
the tight-binding band of the flux-free case, and the energy versus 
$0 \le \alpha \le 1$ diagram reveals a self-similar fractal structure that
is often called the Hofstadter butterfly. 
Since its original proposal back in 1976~\cite{hofstadter76}, 
even though this simple model have continuously sustained the 
attention of a broad range of physicists, and despite all the previous 
efforts, its intricate butterfly spectrum remains to be observed and 
utilized in a clear-cut fashion~\cite{kuhl98, dean13, ponomarenko13}.

One of the historical drawbacks has been the competing length scales. 
This is because $\alpha$ is a direct measure of the ratio of the lattice 
spacing and the cyclotron radius of the charge carrier, 
leading to $\alpha \ll 1$ for typical electronic 
crystals even for the largest magnetic field ($\sim 100$ Tesla) that is 
attainable in a laboratory. Thus, in order to access a sizeable range 
of $\alpha$, one needs to have either an artificial lattice potential 
or an artificial magnetic field with high controllability.
For instance, with the advent of artificial gauge fields in cold-atom 
experiments~\cite{dalibard11}, $\alpha$ may be tuned at will in 
laser-generated optical lattices, which is one of the thriving themes 
in modern atomic and molecular physics. 
The current work is about the time-reversal-symmetric version of this 
model which describes effectively a two-flavored quantum particle, 
e.g., a spin-$1/2$ fermion, with opposite magnetic fields for 
its flavors~\cite{hofstetter12}. 
While such a setup sounds like a bizarre scenario for electronic 
systems, it has already been realized with cold 
atoms~\cite{ketterle13trs, bloch13trs}. 

Motivated by this success~\cite{troyer14, torma15, umucalilar17}, 
here we take advantage of the time-reversal symmetry that is 
manifested by the Hofstadter-Hubbard model with opposite 
magnetic fields, and develop a BCS mean-field theory for 
its low magnetic-flux regime. 
Given the simplicity of the spatially-uniform SF phase in this model, 
we believe it offers an ideal platform for studying the interplay of 
interactions and quantum-Hall physics, e.g, the topological origin 
of the SHI lobes themselves, and the topological SHI-SF phase 
transitions are some of its notable outcomes. 
For instance, we show in this paper that the butterfly spectrum plays 
a major role in the weakly-interacting regime such that the SF phase 
exhibits a number of unambiguous characteristic features both 
in the ground state and at finite temperatures. 
In the strongly-interacting regime, however, this model reduces 
effectively to the usual Hubbard model without any additional physics. 
This is simply because once a Cooper pair is formed, its 
center-of-mass kinematics is neutral against the time-reversal 
symmetric magnetic flux. 

The rest of this paper is organized as follows. 
We first introduce the time-reversal-symmetric Hofstadter model 
in Sec.~\ref{sec:hhm}, and then derive a set of self-consistency 
equations in Sec.~\ref{sec:sce} for the BCS mean-field. 
The analysis of these equations is presented in Sec.~\ref{sec:gs}
for the ground state, where we discuss the SF order parameter in 
Sec.~\ref{sec:op}, SHI-SF transition boundary in Sec.~\ref{sec:shisf}, 
condensate fraction in Sec.~\ref{sec:condpar}, and SF fraction
in Sec.~\ref{sec:supden}.
Furthermore, a similar analysis is presented in Sec.~\ref{sec:criticaltemp}
for finite temperatures, where we discuss the critical BCS transition 
temperature in Sec.~\ref{sec:bcs} and the critical BKT transition 
temperature in Sec.~\ref{sec:bkt}. The paper ends with a brief summary 
of our conclusions in Sec.~\ref{sec:conc}.

\section{Mean-Field Theory}
\label{sec:mft}

In this section, we first introduce the parameters of the model Hamiltonian, 
and then derive a set of self-consistency equations that is based on the 
mean-field decoupling approximation for the BCS pairs.

\subsection{Hofstadter model with $\mathcal{T}$ symmetry}
\label{sec:hhm}

It turns out that the Hofstadter butterfly spectrum is not only symmetric 
around zero energy due to the particle-hole symmetry of the parent 
Hamiltonian but it also has a mirror symmetry around $\alpha = 0.5$. 
Thus, $\alpha = 0.5$ corresponds to the maximally attainable magnetic 
flux within the model in such a way that the flux-free $\alpha = 0$ system
is identical to $\alpha = 1$~\cite{hofstadter76}.  In order to restore 
the time-reversal symmetry into the Hamiltonian, one considers a 
pseudo-spin-$1/2$ fermion experiencing an opposite magnetic flux 
for the components with an equal magnitude, 
i.e., $\alpha_\uparrow = -\alpha_\downarrow$
~\cite{troyer14, torma15, hofstetter12, umucalilar17}.

Given that the energy spectrum of the flux-free 
(to be more precise either $q = \infty$ or $q = 1$) system
$
\varepsilon_\mathbf{k} = -2t \cos(k_xa) - 2t \cos(k_ya)
$ 
is restricted to an energy window of $8t$, where $\mathbf{k} \equiv (k_x, k_y)$ 
is the crystal momentum and $a$ is the lattice spacing, and given the 
multi-band structure of the butterfly for any finite $q$, the widths of the 
bands get progressively narrower with increasing $q$ from 2. 
In turns out that the $n$th energy band is exponentially localized 
in energy near $\varepsilon_n$.
This leads eventually to an infinite set of nearly-flat butterfly bands 
in the $q \to \infty$ limit, recovering the quantum-Hall regime of 
discrete Landau levels. 
Note the striking structural difference between the spectrum of the 
precise $q = \infty$ case and that of the $q \to \infty$ limit.
Next we focus our previous analysis~\cite{umucalilar17} 
of the time-reversal-symmetric Hofstadter-Hubbard model to the 
large-$q$ regime, and explore its SF phase transition within the BCS 
mean-field pairing approximation.

\subsection{Self-consistency equations}
\label{sec:sce}

In the nearly-flat butterfly-bands regime, where the band widths are 
assumed to be much smaller than the band gaps in such a way that 
$\varepsilon_{\mathbf{k}n} \to \varepsilon_n$ for every $\mathbf{k}$ 
state in all bands that are indexed here by $n = 0,1,\dots,q-1$ 
in the first magnetic Brillouin zone, the mean-field Hamiltonian $H$ 
per site can be written as
\begin{eqnarray}
\frac{H}{M} = \frac{1}{q} \sum_{n \sigma} \xi_n d^\dag_{n\sigma} d_{n\sigma}
- \frac{\Delta}{q} \sum_n \left( d^\dag_{n\uparrow} d^\dag_{n\downarrow} +\text{H.c.} \right )
- \frac{\Delta^2}{U}.
\label{eqn:ham}
\end{eqnarray}
Here, $M$ is the number of lattice sites, $\xi_n=\varepsilon_n - \mu$ is the 
common flat-dispersion relation $\varepsilon_n$ shifted by the common chemical 
potential $\mu$, $d^\dag_{n\sigma}$ ($d_{n\sigma}$) creates (annihilates) 
a $\sigma$ fermion in band $n$, and
$
\Delta = (|U|/q)\sum_n \langle d_{n\downarrow} d_{n\uparrow} \rangle
$ 
is the order parameter characterizing a spatially-uniform SF phase of 
$\mathcal{T}$-symmetric Cooper pairs that are made of $|n \uparrow, +\mathbf{k}\rangle$
and $|n \downarrow, -\mathbf{k} \rangle$ fermions with a stationary 
center of mass momentum.
In addition, $U \le 0$ is the onsite interaction between $\uparrow$ 
and $\downarrow$ fermions, $\langle \ldots \rangle$ denotes the thermal 
average, $\text{H.c.}$ is the Hermitian conjugate, and $\Delta$ is assumed 
to be real without losing generality~\cite{umucalilar17}.
This is unlike the usual Hofstadter-Hubbard model with a broken $\mathcal{T}$ 
symmetry, where various competing SF phases require nontrivial sets 
of $q \times q$ order parameters, e.g., a vortex-lattice 
solution~\cite{umucalilar16}. 

Thanks to the quadratic dependence on the fermion operators, it is a straightforward 
task to solve for the thermal properties of Eq.~(\ref{eqn:ham}) via the minimization
of the resultant thermodynamic potential. For instance, a self-consistent solution 
for $\Delta$ and $\mu$ can be obtained by solving the order parameter and 
number equations~\cite{typo}, 
\begin{eqnarray}
\label{eqn:opeqn}
\frac{1}{|U|} &=& \sum_n \frac{1}{2qE_n} \tanh\left(\frac{E_n}{2T}\right), \\
F &=& 1 - \sum_n \frac{\xi_n}{qE_n} \tanh\left(\frac{E_n}{2T}\right),
\label{eqn:numeqn}
\end{eqnarray}
for a given set of parameters. Here, the total particle filling $0 \le F = N/M \le 2$ 
per site is determined by
$
F = (1/q) \sum_{n\sigma} \langle d_{n\sigma}^\dag d_{n\sigma} \rangle,
$ 
$E_n = \sqrt{\xi_n^2 + \Delta^2}$
is the energy spectrum for the quasiparticles in band $n$, $k_B \to 1$ is the 
Boltzmann constant, and $T$ is the temperature. Note that the aforementioned 
particle-hole symmetry of the Hamiltonian manifests in Eqs.~(\ref{eqn:opeqn}) 
and~(\ref{eqn:numeqn}) around $\mu = 0$ or equivalently half filling $F = 1$.

When the strong-coupling condition $\Delta \gg t$ or equivalently $|U| \gg t$ 
is satisfied for all of the butterfly bands, we recover the familiar expressions 
of a flux-free system, e.g., 
$
\Delta_0 = (|U|/2-4t^2/|U|) \sqrt{F(2-F)}
$
and
$
\mu = -(|U|/2-8t^2/|U|) (1-F)
$
are both proportional to the binding energy ($\propto |U|$) of the Cooper pairs 
at $T = 0$, leading eventually to $\Delta_0 = (|U|/2) \sqrt{F(2-F)}$ 
and $\mu = -(|U|/2) (1-F)$ in the strict limit of structureless molecules.
This is not surprising because once a Cooper pair is formed, its
center-of-mass kinematics is neutral against the time-reversal-symmetric 
magnetic flux~\cite{umucalilar17}. 
Alternatively, the only way a tightly-bound molecule to move from a site 
$i$ to $j$ in lattice models is via the virtual ionisation of its constituents, 
and given the binding energy $|U|$ as the penalty-cost for breaking a pair,
this leads to an effective hopping parameter
$
t_{ijm} = 2 t_{ij \uparrow} t_{ij\downarrow}/|U|
$
for the molecules~\cite{iskin08}. Therefore, the effective hopping amplitude and 
magnetic flux of the molecules can be written as $t_m = 2t^2/|U|$ and 
$\alpha_m = \alpha_\uparrow + \alpha_\downarrow = 0$, respectively. 
Furthermore, these Cooper pairs are intrinsically hardcore by their 
composite nature as dictated by the Pauli exclusion principle in the 
$|U|/t \to \infty$ limit, and each site is either empty or singly occupied 
by one of them. These two local states may well be represented as 
an SU(2) algebra of the effective spin, and it turns out that the BCS 
mean-field corresponds precisely to the classical approximation for the 
effective spin model in the molecular limit.

On the other hand, when $\mu = \varepsilon_{\bar n}$ coincides with one 
of the nearly-flat butterfly bands at $U = 0$, the strong-coupling condition 
$\Delta \gg t_{\bar n}^{eff}$ may immediately be satisfied for that particular 
band with a nonzero $U$ including the $U \to 0$ limit. 
This is because since the effective width of all of the bands $t_n^{eff}$ 
decreases dramatically from $t$ with increasing $q$ 
in the $q \gg 1$ regime, the strong-coupling condition 
$|U| \gg t_{\bar n}^{eff}$ can always be achieved for the $\bar{n}$ band 
with increasing $q$ no matter what $U$ is. However, this is not the case for the 
rest of the bands as they are energetically separated from the $\bar{n}$ band 
by varying single-particle gaps leading to $|\xi_{n \ne {\bar n}}| \gg \Delta$ 
when $\Delta \to 0$. 

Next we reveal a number of unusual properties of the model Hamiltonian 
given in Eq.~(\ref{eqn:ham}) by solving Eqs.~(\ref{eqn:opeqn}) 
and~(\ref{eqn:numeqn}) first in the ground state and then at finite temperatures.

\section{Analysis of the Ground State}
\label{sec:gs}

After setting $T = 0$ in this section, we first discuss the characteristic features 
of the SF ground state near the SHI-SF transition boundary by a thorough 
analysis of the order parameter, and then explore its peculiar effects on 
the condensate and SF fractions.

\subsection{Order parameter}
\label{sec:op}

The self-consistency Eqs.~(\ref{eqn:opeqn}) and~(\ref{eqn:numeqn}) for 
the ground-state $\Delta_0$ and $\mu$, i.e.,
$
1/|U| = \sum_n 1/(2qE_n)
$
and
$
F = 1 - \sum_n \xi_n/(qE_n),
$
are analytically tractable in two generic cases.
For instance, when $\mu = \varepsilon_{\bar n}$ coincides with one of the 
nearly-flat butterfly bands at $U = 0$, we find
$
\Delta_0 = (|U|/2)\sqrt{F_{\bar n}(2/q-F_{\bar n})}
$
and
$
\mu_0 = \varepsilon_{\bar n} + (|U|/2) (F_{\bar n}-1/q)
$
for the $U \to 0$ limit~\cite{note1}. We note that the total filling is given by
$
F = 2{\bar n}/q + F_{\bar n}
$
where
$
0 \le F_{\bar n}  = 1/q - \xi_{\bar n}/(qE_{\bar n}) \le 2/q
$ 
is the filling of the ${\bar n}$th band, and that the particle-hole symmetry 
of the Hamiltonian manifests in the ${\bar n}$th band around 
its half filling $F_{\bar n} = 1/q$ or equivalently $\mu = \varepsilon_{\bar n}$.
Thus, unlike the usual BCS superfluids where $\Delta_0$ grows exponentially 
with $|U|$, here the growth is linear and much faster. As discussed towards 
the end of Sec.~\ref{sec:sce}, while such a linear growth is quite typical for 
BCS superfluids in the strong-coupling or molecular limit, here it arises 
immediately with $U \ne 0$ due to the significant enhancement of the 
single-particle density of states in a nearly-flat band. In addition, we find
$
\Delta = \Delta_0 (1 - 2e^{-\Delta_0/T})
$
as $T \to 0$.

\begin{figure}[htbp]
\includegraphics[scale=0.45]{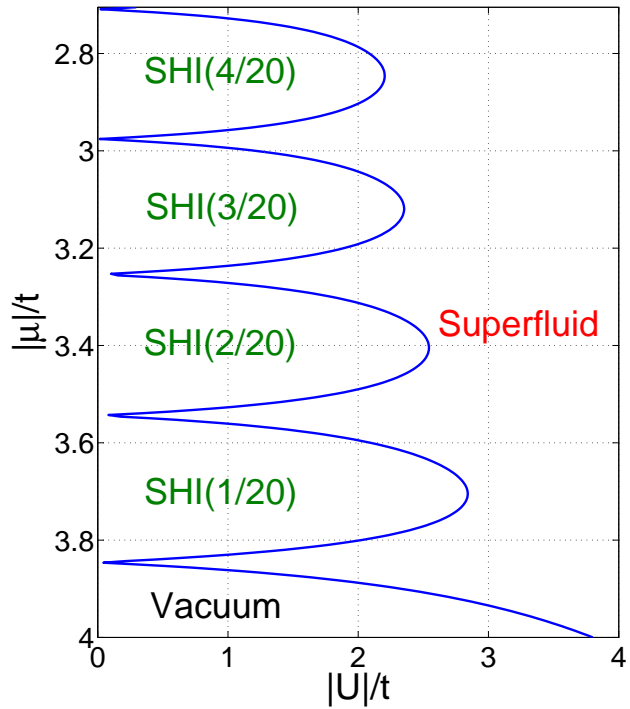}
\caption{(Color online) 
The low-temperature ($T = 10^{-4}t$) phase diagram is shown for 
$\alpha = 1/40$, whose SHI lobes resemble closely to the well-known 
MI lobes of a lattice Bose gas. Here, the total filling $F$ of each lobe 
is indicated in parentheses. 
\label{fig:p1q40}
}
\end{figure}

On the other hand, when $\varepsilon_{\bar{n}-1} < \mu < \varepsilon_{\bar n}$ 
resides within one of the butterfly band gaps at $U = 0$, we find
$
\Delta_0 = \sqrt{(1/U-1/U_c)/C_0}
$
and
$
F = 2{\bar n}/q
$
for the $U \to U_c$ limit, where 
$
1/U_c = -\sum_n 1/(2q|\xi_n|)
$
and 
$
C_0 = \sum_n 1/(4q|\xi_n|^3).
$
Thus, there exists a critical interaction threshold $U_c$ for the SF phase transition 
when there is an energy gap for the single-particle excitations. 
This is illustrated in Fig.~\ref{fig:p1q40} for $\alpha = 1/40$.
Since the butterfly spectrum exhibits a self-similar structure with varying 
$\alpha$, the dependence of $U_c$ on $\mu$, $p$ or $q$ is an intricate 
one~\cite{umucalilar17}. 

In particular, our main focus here is the SHI-SF transition boundary, 
e.g., see Fig.~\ref{fig:p1q40}, which has the familiar shape of insulator 
lobes in the $\mu$ versus $|U|$ plane~\cite{umucalilar17}. 
Note in Fig.~\ref{fig:p1q40} that $U_c$ is small but finite even for 
$\mu = \varepsilon_n$ values due to thermal effects.
In order to gain some physical insight, next we obtain an approximate 
analytical expression for the tip of these SHI lobes when $q \gg 1$.

\subsection{SHI-SF transition}
\label{sec:shisf}

First of all, the expression
$
1/U_c = -\sum_n 1/(2q|\xi_n|)
$
directly suggests that $U_c \to 0$ when $\mu = \varepsilon_{\bar n}$ 
coincides with one of the nearly-flat butterfly bands at $U = 0$, and that 
$U_c \approx - 2q |\xi_{\bar n}|$ when $\mu \to \varepsilon_{\bar n}$ 
from both below or above. In other words, the two consecutive SHI lobes, 
say ${\bar n}$th lobe with filling $2({\bar n}-1)/q$ and $({\bar n}+1)$th lobe 
with filling $2{\bar n}/q$, touch each other precisely 
at $\mu = \varepsilon_{\bar n}$ when $U_c = 0$. 
This leads to a symmetric and linear transition boundary in the 
$\mu$ versus $|U|$ plane, where the hole branch of the ${\bar n}$th lobe 
and the particle branch of the $({\bar n}+1)$th lobe have exactly 
opposite slopes $d \mu/d|U_c| = \pm 1/(2q)$ depending only on $q$. 
When $\alpha = 1/q$, since increasing $q$ rapidly flattens the lowest 
butterfly bands, the analysis given above is quite generic and applies even 
to small $q \gtrsim 8$ values~\cite{umucalilar17}.
When $\mu = 0$ and $q$ is even, the physics is quite different~\cite{sm-sf}.

Furthermore, it is possible to make analytical progress for the tip of the SHI 
lobes in the quantum-Hall regime of discrete Landau levels based on the 
following analogy. Plugging the maximal density 
$
\rho_{max} = m \omega_B/(\hbar \pi)
$ 
per Landau level into the maximal filling 
$
F_{max} = 2/q = \rho_{max} a^2
$
per butterfly band, and using
$
m = \hbar^2/(2ta^2)
$ 
for the effective mass of the particles in the $q \to \infty$ limit, we identify the 
band gap between the Landau levels as 
$
A = \hbar \omega_B = 4\pi t/q.
$
Here, $\omega_B = q_0 B/m$ is the cyclotron frequency of a charged particle 
with mass $m$ and charge $q_0$ in an external 
$\mathbf{B} = B \mathbf{\widehat{z}}$ field.
While this expression is in perfect agreement with the first-order 
(in the small parameter $1/q$) perturbative approach for the lowest butterfly 
band gap when $\alpha = 1/q$~\cite{roy14}, the higher band gaps gradually 
deviate from this value for any finite $q$. In fact, it otherwise would give 
a uniform gap of order $8t/q$, given that the total band width $8t$ is split
into $q$ nearly-flat butterfly bands. 
See Fig.~\ref{fig:p1q40} where $q = 40$.

For the $q \gg 1$ regime, noting that the $n$th energy band is exponentially 
localized in energy near $\varepsilon_n$, here we assume 
$
\varepsilon_n \approx -4t + A(n+1/2),
$ 
where the nearly-flat butterfly bands are again indexed by $n = 0,\ldots,q-1$. 
We note that since the exact band widths are not given by $A$ for finite $q$, 
this assumption clearly overestimates $|U_c|$ for all of the SHI lobes, 
and more importantly, it rapidly fails for the higher SHI lobes with 
increasing ${\bar n}$.
Under this approximation, by plugging $\mu = -4t + As$ into the expression 
for $U_c$ to evaluate the tip of the $s = 1,2,\ldots$th SHI lobe, and using
$
\sum_{j=1}^J 1/(2j-1) \approx [\ln(4J)+\gamma]/2 
$
for large $J$ with $\gamma \approx 0.577$ the Euler's constant, we obtain
\begin{equation}
|U_c^{(s)}| = \frac{2qA}{\gamma+\ln(4q-4s)+\sum_{j=1}^s \frac{2}{2j-1}},
\label{eqn:Uc}
\end{equation}
leading to $|U_c^{(1)}| \approx 8\pi t/(3.9635+\ln q)$ for the first lobe.
In Table~I, we compare this simple expression with that of the fully-numerical 
calculation that is based on the exact butterfly spectrum. Note that 4 
of the lowest SHI lobes are illustrated in Fig.~\ref{fig:p1q40} for $q = 40$.
While it is uplifting to see that the relative accuracy of the analytical 
expression improves monotonously with increasing $q$, the logarithmic 
convergence makes this progress very slow. Given the need 
for days-long computation times even for $q \sim 1000$, Eq.~(\ref{eqn:Uc}) 
may still offer an enormously convenient way for a reasonable estimate 
of $U_c$ in the large-$q$ regime~\cite{note2}.

\begin{table}[htbp]
\label{table:Uc}
\begin{tabular}{c|ccc}
$q$   & $|U_c^{(1)}|/t$ & $|U_c|/t$ & Error ($\%$) \\
\hline
40     &  3.284  &  2.841  &  15.6 \\
100   &  2.933  &  2.614  &  12.0 \\
1000 &  2.312  &  2.126  &  8.60 \\
2000 &  2.173  &  2.009  &  8.15 \\
5000 &  2.014  &  1.873  &  7.50 \\
\end{tabular}
\caption{Comparison of the tip of the first SHI lobe shows that the 
relative convergence between the analytical approximation given in 
Eq.~(\ref{eqn:Uc}) and the numerical result that is based on the 
exact butterfly spectrum is logarithmic.}
\end{table}
\subsection{Condensate fraction}
\label{sec:condpar}

The onsite pairing parameter 
$
\Psi_i = \langle c_{i\uparrow} c_{i \downarrow} \rangle,
$
where $c_{i\sigma}$ annihilates a $\sigma$ fermion on site $i$, 
characterizes the number of Cooper pairs in the usual way, and it is 
directly related to the onsite SF order parameter $\Delta_i$ via 
$\Delta_i = U\Psi_i$. Since $\Delta_i$ is uniform in space in this paper,
this leads to
$
\Delta = (|U|/M) \sum_{n\mathbf{k}} \Psi_{\mathbf{k}n}
$
with the pair wave function 
$
\Psi_{\mathbf{k}n} = [\Delta/(2E_{\mathbf{k}n})] \tanh[E_{\mathbf{k}n}/(2T)]
$
for a given $\mathbf{k}$ state in the first magnetic Brillouin zone. 
Thus, we identify
$
N_c = 2\sum_{n\mathbf{k}} |\Psi_{\mathbf{k}n}|^2
$
as the total number of condensed particles in general~\cite{iskin08}, 
leading to 
$
F_c = N_c/M = (2/q) \sum_n |\Psi_n|^2
$ 
as their filling in the nearly-flat butterfly-bands regime where 
$\Psi_{\mathbf{k}n} \to \Psi_n$ for every $\mathbf{k}$ state of 
all butterfly bands. 

The condensate filling
$
F_c = [\Delta_0^2/(2q)] \sum_n 1/(\xi_n^2 + \Delta_0^2)
$
for the ground state is also analytically tractable in two generic cases.
When $\mu = \varepsilon_{\bar n}$ coincides with one of the 
nearly-flat butterfly bands at $U = 0$, we find
$
F_c \approx F_{{\bar n}c} = F_{\bar n}(2- qF_{\bar n})/2
$
for the $U \to 0$ limit as long as $\Delta_0 \ne 0$. Recall that the total 
particle filling is $F = 2{\bar n}/q + F_{\bar n}$ for this limit. Thus, $F_{{\bar n}c}$ 
is independent of $U$, and while all of the particles or holes in band ${\bar n}$ 
are condensed at its low- and full-filling regimes, where 
$F_{{\bar n}c} \to F_{\bar n}$ and $F_{{\bar n}c} \to 2/q-F_{\bar n}$, 
respectively, only a half of the particles is condensed at half filling
where $F_{{\bar n}c} = 1/(2q)$. This is simply because, since the particles 
are strongly interacting in the $\bar{n}$ band as long as $\Delta_0 \ne 0$ 
no matter how small it is, the fraction of condensed pairs in band ${\bar n}$ 
achieves its maximal value immediately with $U \ne 0$. This, however, 
is not the case for the rest of the particles in the lower bands as 
$\Delta_0 \ll |\xi_{n \ne \bar{n}}|$ when $U \to 0$.

On the other hand, when $\varepsilon_{\bar{n}-1} < \mu < \varepsilon_{\bar n}$ 
resides within one of the butterfly band gaps at $U = 0$, we find
$
F_c = (C_1/C_0) (1/U-1/U_c)
$
for the $U \to U_c$ limit where $\Delta_0 \ll |\xi_n|$~\cite{note3}.
Here,
$
C_0 = \sum_n 1/(4q|\xi_n|^3)
$
and
$
C_1 = \sum_n 1/(2q|\xi_n|^2).
$
Recall that the total particle filling is $F = 2{\bar n}/q$ for this limit.

Thus, unlike the usual BCS superfluids where 
$
F_c \propto \Delta_0/t
$
grows exponentially slow with $|U|$~\cite{iskin08}, 
here $F_c$ either immediately 
saturates or grows linearly with $|U|$, depending on where $\mu$ lies 
at $U = 0$. This peculiar $U$ dependence is a direct reflection of 
that of $\Delta_0$ discussed in Sec.~\ref{sec:op}, and its effects 
are naturally not limited to the condensate fraction. For instance, 
next we show that $\Delta_0$ leaves its definitive signatures 
on the SF fraction as well.

\subsection{Superfluid fraction}
\label{sec:supden}

By definition, the SF density is directly proportional to the familiar 
phase stiffness of the effective phase-only Hamiltonian, 
e.g., see Sec.~\ref{sec:bkt} for a brief discussion about the critical 
BKT transition temperature. 
For instance, when $\mu = \varepsilon_{\bar n}$ coincides 
with one of the nearly-flat butterfly bands at $U = 0$, 
the phase stiffness is given by
$
\Gamma = (2\bar{n} + 1) q \Delta^2/(2\pi |U|)
$
for the $U \to 0$ limit~\cite{torma15}, and it reduces to
$
\Gamma_0 = (2\bar{n} + 1) F_{\bar n}(2-qF_{\bar n}) |U| / (8\pi)
$ 
at $T = 0$. Substituting $t = \hbar^2/(2ma^2)$ for the effective mass of 
the particles in the $q \to \infty$ limit and $\rho_s = F_s/a^2$ for the SF 
density into the relation 
$
\Gamma = \hbar^2 \rho_s/(4m)
$
shown in Sec.~\ref{sec:bkt}, we obtain the SF fraction as
$
F_s/F = (2\bar{n} + 1) qF_{\bar n}(2-qF_{\bar n}) |U|
/ [4\pi t (2\bar{n} + qF_{\bar n}) ].
$
While this fraction becomes
$
F_s/F = (2-qF_0) |U| / (4\pi t)
$
for the $\bar{n} = 0$ band with any given filling $0 \le F_0 \le 2/q$, 
it becomes
$
F_s/F = |U| / (4\pi t)
$
for any ${\bar n}$ band near their half fillings $F_{\bar n} = 1/q$.
When $q$ is even, the latter $q$-independent expression may also 
be reproduced for the central bands by taking the $\mu = 0$ or 
equivalently $F = 1$ limit in the large-$q$ regime.
Since the low-energy dispersion involves $q$ Dirac cones, it can be 
shown that
$
\Gamma_0 = q \sqrt{\Delta_0^2+\mu^2} /(8\pi)
+ q\Delta_0^2 \ln [(|\mu| + \sqrt{\Delta_0^2+\mu^2})/\Delta_0] /(8\pi |\mu|),
$
for any $q$ in the weak-coupling limit~\cite{kopnin10}. 
This reduces to $\Gamma_0 = q\Delta_0/(4\pi)$ as $\mu \to 0$, 
leading eventually to
$
F_s = |U| / (4\pi t)
$
for the flattened centrally-symmetric bands where $\Delta_0 = |U|/(2q)$ 
at their combined half filling. 

Thus, unlike the usual BCS superfluids where the ground state of 
an entire Fermi gas turns from normal to SF immediately with 
$U \ne 0$ at $T = 0$, i.e., $F_s = F$ in the $U \to 0$ limit, 
here the SF density sets in gradually with a linear growth in $|U|$. 
Having characterized the SF ground state near the SHI-SF transition 
boundary, next we discuss the SF phase at finite temperatures.

\section{Critical Temperatures}
\label{sec:criticaltemp}

After setting $T \ne 0$ and $\Delta \to 0$ in this section, we first 
discuss the characteristic features of the SF phase near the SF-Normal 
transition boundary by a thorough analysis of the critical BCS 
temperature, and then compare it with the critical BKT temperature.

\subsection{BCS transition temperature}
\label{sec:bcs}

The self-consistency Eqs.~(\ref{eqn:opeqn}) and~(\ref{eqn:numeqn}) 
for the critical BCS transition temperature $T_c$ and $\mu$, i.e.,
$
1/|U| = \sum_n \tanh[\xi_n/(2T_c)]/(2q\xi_n)
$
and
$
F = 1 - \sum_n \tanh[\xi_n/(2qT_c)],
$
are again analytically tractable in two generic cases.
For instance, when $\mu = \varepsilon_{\bar n}$ coincides with one 
of the nearly-flat butterfly bands at $U = 0$, while we find
$
T_c = |U|/(4q) 
$
and
$
\mu = \varepsilon_{\bar n} + 2T (qF_{\bar n} - 1)
$
near its half filling $F_{\bar n} = 1/q$ for the $U \to 0$ limit, we find
$
T_c = -|U|/[2q\ln(q F_{\bar n}/2)]
$
in its low-filling $q F_{\bar n} \to 0$ regime and
$
T_c = -|U|/[2q\ln(1 - q F_{\bar n}/2)]
$
in its full-filling $q F_{\bar n} \to 2$ regime.
Note that while 
$
T_c/\Delta_0 = 0.5
$
is close to the well-known BCS ratio
$
T_c/\Delta_0 = e^\gamma/\pi \approx 0.567
$
near half filling, it diverges as $\lim_{x \to 0^+} -1/(\sqrt{x}\ln x)$ in both 
of the low- and full-filling regimes. The latter is caused by the depletion 
of the low-energy density of states near the band edges, having more
dramatic effects on the ground-state properties than the thermal ones. 
In addition, we find
$
\Delta = \sqrt{12} T_c \sqrt{1-T/T_c}
$
for the $T \to T_c$ limit near half filling, leading to
$
\Delta = \sqrt{3} \Delta_0 \sqrt{1-T/T_c}.
$
It is not only the precise form of this latter expression that is identical 
to that of the usual BCS one but the prefactor $\sqrt{3} \approx 1.732$ 
also coincides with $e^\gamma \sqrt{8/[7\xi(3)]} \approx 1.736$, 
where $\xi(3) \approx 1.202$ is the Riemann-zeta function. While we 
are delighted by this almost perfect coincidence, it is probably an 
accidental match without any deep physical reasoning.

On the other hand, when $\varepsilon_{\bar{n}-1} < \mu < \varepsilon_{\bar n}$ 
resides within one of the butterfly band gaps at $U = 0$, we find
$
\Delta = [\sum_n e^{-|\xi_n|/T_c}/(qC_0T_c)]^{1/2} \sqrt{1-T/T_c}
$
for the $T \to T_c$ limit, leading to
$
\Delta = \sqrt{U_c^2 F/[2T_c (U_c-U)]} \Delta_0 \sqrt{1-T/T_c}
$
in the $F \to 0$ regime, and
$
\Delta = \sqrt{U_c^2 (2-F)/[2T_c (U_c-U)]} \Delta_0 \sqrt{1-T/T_c}
$
in the $F \to 2$ regime. Thus, similar to the usual BCS superfluids near 
$T_c$, here $\Delta$ again decays as $\sqrt{1-T/T_c}$, which is one 
of the characteristic signatures of a second-order phase transition, but
with a larger coefficient as $U \to U_c$. The large prefactor is again 
caused by the depleted density of states within the band gap as 
$\Delta_0$ is expected to be more susceptible to such changes than $T_c$. 

All of these finite-$T$ results are based on mean-field theory, the 
equivalent of which is known to capture, quite accurately, 
the low-$T$ properties of a 3D continuum Fermi gas in the 
weak-coupling BCS and strong-coupling BEC limits. 
In fact, this success has been the main motivation for its
extensive use for the entire range of interactions, even though its 
accuracy is also known to be much less in the so-called BCS-BEC 
crossover. However, since the model at hand is a 2D lattice one with a 
complicated multi-flat-band energy spectrum, next we discuss the critical 
BKT temperature as a means to attest the validity of our finite-$T$ results.

\subsection{BKT transition temperature}
\label{sec:bkt}

The critical BKT transition temperature $T_{BKT}$ for a 2D XY model 
is given by the universal relation
$
T_{BKT} = \pi \Gamma/2,
$
where $\Gamma$ is the phase stiffness that is defined via the effective 
phase-only Hamiltonian
$
H_{XY} = (\Gamma/2) \int d^2\mathbf{r} (\nabla \theta_\mathbf{r})^2,
$
under the assumption that the SF order parameter 
$
\Delta_\mathbf{r} = \Delta e^{i\theta_\mathbf{r}}
$
has a spatially-varying phase.
Setting $\theta_\mathbf{r} = \mathbf{K} \cdot \mathbf{r}$ for a uniform
condensate density with $\hbar \mathbf{K}$ the pair momentum,
we note that
$
H_{XY}/A = \rho_s \hbar^2 K^2 / (4m) = \rho_{sm} v^2/2,
$
where $A$ is the area, $\mathbf{v} = \hbar \mathbf{K}/(2m)$ is the velocity 
of the pairs, and $\rho_{sm} = m \rho_s$ is the SF-mass density.
Thus, a self-consistent solution for $\Delta$, $\mu$ and $T_{BKT}$ can 
be obtained by solving the order parameter and number equations given
in Eqs.~(\ref{eqn:opeqn}) and~(\ref{eqn:numeqn}) together with 
the universal relation. 

While the calculation of $\Gamma$ is typically a highly-nontrivial task 
for an arbitrary $q$ value, it has recently been performed for the 
large-$q$ regime of our time-reversal-symmetric model~\cite{torma15}. 
For instance, when $\mu = \varepsilon_{\bar n}$ coincides with one 
of the nearly-flat butterfly bands at $U = 0$, such that 
$
\{ |\xi_{\bar n}|, T_{BKT} \} \ll |\xi_{n \ne \bar{n}}|,
$ 
one finds
\begin{eqnarray}
\frac{1}{|U|} &=& \frac{1}{2qE_{\bar n}} \tanh\left(\frac{E_{\bar n}}{2T_{BKT}}\right), \\
F_{\bar n} &=& \frac{1}{q} - \frac{\xi_{\bar n}}{qE_{\bar n}} \tanh\left(\frac{E_{\bar n}}{2T_{BKT}}\right), \\
T_{BKT} &=& \frac{(2 {\bar n} +1)q }{4|U|} \Delta^2,
\label{eqn:bkt}
\end{eqnarray}
for the $U \to 0$ limit. Here, the total particle filling is given by 
$F = 2\bar{n}/q + F_{\bar n}$, and this set of equations naturally gives
$T_{BKT} \le T_c$ given that $T_c$ is determined by the $\Delta \to 0$ 
condition. 

We note that the physical origins of $T_c$ and $T_{BKT}$ 
are quite different. While $T_c$ has to do with the BCS pairing of particles 
or where $\mu$ lies within any of the given nearly-flat butterfly bands, 
$T_{BKT}$ has to do with the phase coherence and the true SF phase 
transition involving the entire Fermi gas. In other words, $T_c$ and $T_{BKT}$, 
respectively, set the scale for the onset of Cooper pairing and the binding of 
vortex-antivortex pairs. This is the physical insight for the explicit ${\bar n}$ 
dependence appearing in Eq.~(\ref{eqn:bkt}).
Next we show that while $T_{BKT} \to T_c$ for ${\bar n} \gtrsim 4$ 
and $T_{BKT} \lesssim T_c$ for ${\bar n} \lesssim 3$ near their half fillings 
$F_{\bar n} = 1/q$, we find $T_{BKT} \ll T_c$ for any $\bar{n}$ 
if its filling $F_{\bar n}$ is sufficiently away from $1/q$.

For instance, assuming $T_{BKT} \approx T_c$, we may substitute the 
analytic $\Delta$ expression that is derived in Sec.~\ref{sec:bcs} for the 
$T \approx T_c$ limit. This leads to
$
T_{BKT}/T_c = (6q {\bar n} + 3q) T_c / [(6q {\bar n} + 3q) T_c + |U| ],
$
showing that the ratio approaches to unity if $T_c \gg |U|/(6q {\bar n} +3q)$.
This last condition can be easily satisfied for large enough ${\bar n}$ around 
its half filling. To illustrate this, we plug the half-filling $T_c$ derived 
in Sec.~\ref{sec:bcs}, and obtain
$
T_{BKT}/T_c = (6 {\bar n} +3) / (6 {\bar n} +7)
$
for any ${\bar n}$. In comparison with the numerical findings $0.75$ and 
$0.82$ for the half-filled ${\bar n} = 2$ and $3$ bands~\cite{torma15}, 
our analytical expression gives $0.79$ and $0.84$.
This good agreement suggests that our expression is quite accurate 
for ${\bar n} \gtrsim 4$ at their half fillings where 
$
T_{BKT} = (6 {\bar n} +3) |U| / (24 q {\bar n} + 28q),
$
and that ${\bar n}$ has to be very large for $T_{BKT} \approx T_c$ 
away from their half fillings.

Similarly, assuming $T_{BKT} \ll T_c$, we may substitute the analytic 
$\Delta_0$ expression that is derived in Sec.~\ref{sec:op} for the 
$T \ll T_c$ limit. This leads to
$
T_{BKT} = (2 {\bar n} +1) F_{\bar n}(2-qF_{\bar n}) |U| /16,
$
which is independent of large-$q$ at $\mu = 0$ giving $T_{BKT} \approx |U|/16$.
Thus, we find $T_{BKT}/T_c = (2 {\bar n} +1)/4$ for half filling, which
coincides exactly with the numerical finding $1/4$ for the 
${\bar n} = 0$~\cite{torma15} band. This perfect agreement may 
further suggest that our expression
$
T_{BKT} = F_0(2-qF_0) |U| / 16
$
for the ${\bar n} = 0$ band may work well for all other fillings as well. 
As ${\bar n}$ increases, $T_{BKT}$ eventually approaches to $T_c$. 
For instance, away from half filling, the ratio
$
T_{BKT}/T_c = (2 {\bar n} +1)q F_{\bar n} \ln [2/(qF_{\bar n})]
$
ultimately vanishes in the $qF_{\bar n} \to 0$ limit for any finite ${\bar n}$.
Thus, we conclude that while
$
T_{BKT} = (2 {\bar n} +1) F_{\bar n} |U| / 8
$
in the low-filling regime, 
$
T_{BKT} = (2 {\bar n} +1) (2/q-F_{\bar n}) |U| / 8
$
in the full-filling one. 

We note in passing that, when the strong-coupling condition $|U| \gg t$ is 
satisfied for all of the butterfly bands, one must recover the familiar 
expressions of a flux-free system, where $T_c \propto |U|$ is proportional 
to the binding energy $|U|$ and $T_{BKT} \propto t^2/|U|$ is proportional
to the effective mass $\hbar^2/(2t_ma^2)$ of the structureless molecules,
leading to $T_{BKT} \ll T_c$ as well. 
Having characterized the SF phase at finite temperatures, 
we end this paper with a brief summary of our conclusions.

\section{Conclusions}
\label{sec:conc}

To summarize, we took advantage of the time-reversal symmetry 
that is manifested by the Hofstadter-Hubbard model with opposite 
magnetic fields, and developed a BCS mean-field theory for its 
multi-band spectrum in the nearly-flat butterfly-bands regime. 
In particular, our detailed analysis that includes 
but is not limited to the order parameter, condensate and superfluid 
fractions, and the critical BCS and BKT transition temperatures 
revealed a number of unusual characteristics for the SF transition both 
in the ground state and at finite temperatures. 

Given the simplicity of the spatially-uniform SF phase in this 
time-reversal-symmetric model, e.g., without any serious complications 
coming from the competing phases such as vortex-like excitations, 
and its direct relevance to Lieb-like lattice models exhibiting flat bands, 
we believe the model offers an ideal platform for studying the interplay 
between the pairing correlations and the topologically-nontrivial 
butterfly bands. For instance, a rich variety of topological phases 
and topological phase transitions have recently became some of the 
central topics in modern condensed-matter physics, and both the 
topological nature of the SHI lobes themselves and the topological 
SHI-SF phase transitions are a few of the standout results. 
The unambiguous signatures that are highlighted in this work not only
guide the way in shaping our intuition behind these competing phases, 
but they also serve as ultimate benchmarks for numerically-exact 
QMC simulations.

\begin{acknowledgments}
The author acknowledges funding from T{\"U}B{\.I}TAK and the BAGEP award 
of the Turkish Science Academy, and the discussions with R. O. Umucal\i lar.
\end{acknowledgments}


\begin{thebibliography}{99}
\bibitem{typo}
These equations are simply reduced from the general expressions given
in Ref.~\cite{umucalilar17}, where the extra factor of $q$ appearing in 
the denominator of Eq.~(3) is a typo there.
\bibitem{note1} In the quantum-Hall regime of discrete Landau levels,
our half-filling expression $\Delta_0 = |U|/(2q)$ recovers precisely 
the continuum model of Ref.~\cite{nishida17}, where they find 
$
\Delta_0 = m |g| \omega_B/(4\pi \hbar)
$
with $m = \hbar^2/(2ta^2))$ and $\hbar \omega_B = 4\pi t/q$ 
for the $g = U a^2 \to 0$ limit.
\bibitem{sm-sf} When $\mu = 0$ and $q$ is even, 
the centrally-symmetric energy bands ($n = q/2$ and $q/2-1$) touch each 
other at zero energy, forming $q$ Dirac cones in the first magnetic Brillouin zone.
We have shown previously that~\cite{umucalilar17} the $|U_c|$ versus 
$\alpha$ phase diagram exhibits an oscillatory pattern repeating itself 
in a self-similar structure, due to the fractal nature of the spectrum.
The oscillations are between the $\alpha$ interval 
$(1/2^+, 1/4^-)$; $(1/4^+, 1/6^-)$; etc.,
because as we approach $1/q$ ratios with large $p$ there are more
and more Dirac cones gradually turning the semi-metal phase into a 
normal metal where $U_c \to 0$.
It is such that, for any given prime number $p$, each oscillation consists 
of $p-1$ data points, which can be written in the form
$
\alpha_{jj'} = p/(2pj + 2j') 
$
where $j = 1,2,\dots$ labels the oscillation between the $\alpha$ interval 
$1/(2j)^+$ and $1/(2j+2)^-$, and $j' = 1,2,\dots,p-1$.
If $p$ is not a prime number, there are clearly fewer data points.
For instance, $p = 9$ has $8-2 = 6$ data points since $9/24 = 3/8$ and 
$9/30 = 3/10$ in the first oscillation.
\bibitem{note2} Equation~(\ref{eqn:Uc}) can be mapped to that of the 
continuum model of Ref.~\cite{nishida17} after setting $U = g/a^2$ 
and $2\pi\alpha = \Phi/\phi_0$, where $\alpha = 1/q$, $\Phi = Ba^2$ is 
the flux per unit cell and $\phi_0 = \hbar/q_0$ is the flux quantum. 
This leads to
$
2m |g_c^{(s)}| = 8\pi / [ \gamma+\ln(8\pi \phi_0/\Phi - 4s)+\sum_{j=1}^s 2/(2j-1) ],
$
for the tip of the $s$th lobe located at $\mu = \hbar \omega_B s$.
Furthermore, if the tip of the first SHI lobe $g_c^{(1)}$ is known then the tips 
of the higher lobes can be generated by
$
g_c^{(s)} \approx 1/\lbrace 1/g_c^{(1)} + \sum_{j = 2}^s m/[2\pi (2j-1)] \rbrace.
$
\bibitem{note3} In comparison, we find that the total condensate filling 
grows as
$
F_c = 1/(2C) (1/U-1/U_c)^2
$
in the $\Delta_0 \to 0$ limit when $\mu = 0$ and $q$ is even. 
Here, $C$ is the slope of the low-energy density of states near the 
Dirac cones, i.e., $C |\varepsilon|$ as $\varepsilon \to 0$, and 
$U_c$ is the critical threshold for the semimetal-SF transition~\cite{umucalilar17}.
\end{thebibliography}
\end{document}